\documentclass[prb,aps,amssymb,twocolumn,superscriptaddress,notitlepage]{revtex4-2}
\usepackage{amsmath}
\usepackage{amssymb}
\usepackage{amsthm}
\usepackage{amsfonts}
\usepackage{listings}
\usepackage{physics}
\usepackage{enumerate}
\usepackage{latexsym}
\usepackage{psfrag}
\usepackage{bm}
\usepackage{graphicx}
\usepackage[caption=false]{subfig}
\usepackage{blkarray}
\usepackage{array}
\usepackage{color}
\usepackage[normalem]{ulem}
\usepackage{hyperref}
\usepackage[dvipsnames]{xcolor}
\hypersetup{
     colorlinks = true,
     linkcolor = magenta,
     citecolor = magenta
}

\usepackage{comment}

\def\beq{\begin{equation}}
\def\eeq{\end{equation}}
\def\bal{\begin{aligned}}
\def\eal{\end{aligned}}

\begin{document}
\title{Anisotropy of the Hydrostatic Stress for Hall Droplets with in-plane Magnetic Field
}
\author{Ian Osborne}
\email{ianro2@illinois.edu}
\affiliation{Department of Physics, University of Illinois at Urbana-Champaign, Urbana IL 61801, USA}
\affiliation{Anthony J.~Leggett Institute for Condensed Matter Theory, University of Illinois at Urbana-Champaign, Urbana IL 61801, USA}

\author{Gustavo M.~Monteiro}
\email{gustavo.monteiro@csi.cuny.edu}
\affiliation{Department of Physics and Astronomy, College of Staten Island, CUNY, Staten Island, NY 10314, USA}

\author{Barry Bradlyn}
\email{bbradlyn@illinois.edu}
\affiliation{Department of Physics, University of Illinois at Urbana-Champaign, Urbana IL 61801, USA}
\affiliation{Anthony J.~Leggett Institute for Condensed Matter Theory, University of Illinois at Urbana-Champaign, Urbana IL 61801, USA}

\date{\today}

\begin{abstract}
We examine the hydrostatic stress of electrons strongly confined in a quasi-2D quantum well in the presence of a strong perpendicular and weak in-plane magnetic field. 
This introduces anisotropy into the stress tensor which is inconsistent with the notion of the quantum Hall droplet as a simple two-dimensional electron fluid. 
We show that the breaking of rotational symmetry is a necessary but not a sufficient condition for an anisotropic ground state stress tensor, and that the anisotropic stress originates primarily from quantum effects. 
We demonstrate  this by decomposing the semiclassical trajectories of electrons in the system onto planes in phase space which must be rotated relative to the plane of the fluid to decouple the Hamiltonian. 
\end{abstract}
\maketitle

\section{Introduction}

The observation of viscosity-dominated flow of electrons in two-dimensional solids has sparked a renewed interest in the study of hydrodynamics~\cite{lucas2018hydrodynamics}. 
For clean electronic systems such as graphene where scattering is dominated by momentum-conserving interactions rather than disorder, hydrodynamic effects lead to the emergence of Pouiselle flow and backflow which can be measured experimentally~\cite{sulpizio2019visualizing,bandurin2016negative}. 
Similarly, the presence of a magnetic field leads to the emergence of nondissipative viscosity coefficients such as the Hall viscosity, which has been extensively studied for quantum Hall systems~\cite{berdyugin2019measuring,delacretaz2017transport,read2009nonabelian,read2011hall,bradlyn2012kubo,abanov2014electromagnetic,avron1995viscosity}.
 Unlike ordinary fluids, transport coefficients in quantum fluids (such as quantum Hall systems) do not arise from scattering and are not described in terms of the mean-free-path. 
Instead, the system gap $E_G$ introduces both a time scale $(\hbar/E_G)$ and a length scale $(\hbar/\sqrt{m E_G})$ to the system such that the hydrodynamic behavior arises when expectation values vary slowly within these scales. This allow us to expand the constitutive relations in a gradient expansion where the expansion parameter is governed by this length scale~\cite{jensen2012towards,banerjee2012constraints}.

The standard approach to hydrodynamics takes as its starting point the isotropy (rotational invariance) of ordinary fluids~\cite{landau1987fluid,Kadanoff1963}. 
However, crystalline solids break rotational symmetry, and the flow of electrons in solids can be sensitive to this anisotropy. 
This has motivated a renewed interest in the theory of anisotropic hydrodynamics~\cite{rao2020hall,Huang2022,gromov2017bimetric,gromov2017investigating,offertaler2019viscoelastic,friedman2023hydrodynamics,qi2023anomalous}. 
For two-dimensional fluids, theoretical work in this area has identified constraints on the stress tensor and viscosity tensor that must be satisfied for fluids with any point group symmetry. 
Furthermore, experimental proposals have been put forth for how anisotropic stress and viscosity can be extracted through heat transport or surface wave measurement~\cite{cook2021viscometry,rao2023resolving}. 
Of particular note, Ref.~\cite{Huang2022} argued that for two-dimensional fluids, the ground state stress tensor in the absence of external sources will be isotropic regardless of any rotational symmetry breaking.

Beyond two dimensions, however, our understanding of electron hydrodynamics is less clear, even at the level of hydrostatics. 
One experimentally-relevant system that illustrates this issue is a confined quasi-two-dimensional electron gas in a magnetic field~\cite{eisensteintiltedfqh,eisensteintiltedinteger,shayegantilted,pantilted,csathytiltsecondll,girvintilt1999}. 
Strictly speaking, all two-dimensional electron gases (2DEGs) in quantum wells are quasi-two-dimensional: at low temperatures, almost all electrons lie in the ground state of the quantum well, which we model as a confining potential $V(z)$ in the $z$-direction. 
When such a system is placed in a magnetic field also aligned along $\mathbf{\hat{z}}$ (out of plane), we can ignore the confining potential and treat the system as effectively two-dimensional. 
However, when the magnetic field has an in-plane (e.g., $x$-) component, the Lorentz force couples motion in plane to motion out of plane. 
This can be achieved in experiments on quantum Hall systems in GaAs quantum wells by, for instance, tilting the sample relative to the external magnetic field. 
At the level of the Hamiltonian, at low temperatures one can still attempt to project out the out-of-plane degrees of freedom; doing so maps the titled field system to an effective two-dimensional system with an anisotropic effective mass tensor, where the anisotropy is a function of the in-plane field~\cite{maan1984combined,halonen1990subband,wang2003spontaneous,papic2013fractional}.

However, it was shown in Ref.~\cite{offertaler2019viscoelastic} that this mapping breaks down at the level of hydrostatics. 
Because the stress tensor of the tilted field quantum Hall (TFQH) droplet is defined in three dimensions, the ground state stress defined from the full Hamiltonian does not correspond to the na\"ive expectation one would have for the stress tensor of an anisotropic two-dimensional system. 
While this was shown mechanically in Ref.~\cite{offertaler2019viscoelastic} for the quantum system at zero temperature, the result is counterintuitive. Furthermore, we expect that in the high-temperature limit the TFQH system should have an isotropic stress tensor similar to a classical collisionless plasma. In this work, we will examine the semiclassical hydrostatics of the tilted field system at both zero and nonzero temperature in order to explore the origin of this breakdown at low temperatures. 
By studying the semiclassical solutions to the equations of motion for the tilted field system, we will show that the quantization of phase space orbits is the origin for the anisotropy in the ground state stress tensor. 
Additionally, we will show how, as temperature is increased, the anisotropy in the average stress vanishes and we recover the isotropic form of the stress that we would expect for a two-dimensional fluid. 

To demonstrate this, we begin in Sec.~\ref{sec:tfqh} by introducing our model for the tilted field system in the quantum Hall regime. 
We review the solution of the quantum problem and the construction of the stress tensor operator and its expectation value. 
Next, in Sec.~\ref{sec:classical} we examine the model semiclassically. 
We show first that classically, at any temperature, an electron gas in a tilted field has an isotropic stress tensor. 
To study how anisotropy emerges at the quantum level, we analyze the classical trajectories of the model. 
We show that the dynamics of electrons breaks up into a slow motion mostly aligned in the plane of confinement, and a fast motion mostly aligned out of the plane of confinement. 
Applying Bohr-Sommerfeld quantization to these classical trajectories, we find that the anisotropy in the model emerges from the anisotropy of the Bohr-Sommerfeld action integrals. 
Finally, we compute the temperature dependence of the anisotropy in the stress tensor and show how the stress crosses over from the low-temperature anisotropic quantum result to the high-temperature isotropic classical result.

\section{Tilted Field Quantum Hall (TFQH) Droplet}\label{sec:tfqh}

Despite the predictive power of strictly two-dimensional models, a realistic Hall droplet exists in three spatial dimensions. 
When the magnetic field is oriented perpendicular to the plane of the material, the dynamics of the in-plane and out-of-plane motion are decoupled. Thus, when the confining potential is strong enough that all particles exist in its lowest energy state, the out-of-plane dynamics can be ignored. 
This dimensional reduction is complicated by the presence of an in-plane magnetic field, which couples the in and out-of-plane motion such that the confining potential can no longer be ignored. 
In this section, we introduce our model and review the momentum continuity equation, which we will use to define the stress tensor.

\subsection{Model}
\label{sec:Model}

To effectively model a system in two spatial dimensions, we consider a quantum Hall droplet strongly confined to the $x-y$ plane by an external potential which we approximate as harmonic. 
The energy scale $\omega_0$ of the potential is taken to be much greater than all other energy scales in the Hamiltonian (we set $\hbar=1$ for convenience). 
The choice of harmonic potential leaves the single-particle Hamiltonian exactly solvable, however a consequence of this choice is the equal spacing of the energy spectrum for excitations of the confining potential.

The Hamiltonian for a strong, harmonically confined, non-interacting electron gas with flat, isotropic spatial metric is 
\begin{align}\label{eq:ManyBodyHam}
    H =  \sum_{i }^N \frac{\eta^{\mu\nu}}{2m}\pi^i_\mu \pi_\nu^i + \frac{1}{2}m \omega_0^2 z_i^2.
\end{align}

Where $\mu,\nu \in \{x,y,z\}$ label spatial dimension, $i,j\in \{1,\dots,N\}$ label the particles, $\pi_\nu^i$ are the (kinetic) momentum operators for the particles, $z_i=x_i^3$ is the component of the position operator perpendicular to the plane of confinement, and 
\begin{align}
    \eta^{\mu\nu} = \delta ^{\mu\nu}
\end{align}
is the flat metric (effective mass tensor, in this case). 
The kinetic momentum is given by $ \pi_\mu = p_\mu + A_\mu$, where $A_\mu$ is the electromagnetic vector potential and we take the charge of the electron to be $-|e|=-1$. 
The commutator between the position operator and the kinetic momentum is canonical
\begin{align}
    [x_i^\mu, \pi_\nu^j]=i \delta_i^j\delta_\nu^\mu,
\end{align}
while the commutators between $\pi_\mu$ operators yields the electromagnetic field strength tensor which reduces in our case to
\begin{align}
    [ \pi_\mu^i,  \pi_\nu^j ] &= - i \delta^{ij} \epsilon_{\mu\nu\rho} B^\rho,
\end{align}
where $\epsilon_{\mu\nu\lambda}$ is the totally antisymmetric Levi-Civita tensor ($\epsilon_{123} = -\epsilon_{213} = 1$ in Cartesian coordinates). 

Because we are interested in physics in the presence of an in-plane magnetic field we ``tilt" the magnetic field in the plane of the droplet, such that $\mathbf{B} =  B^z\mathbf{\hat z} + B^x \mathbf{\hat {x}}$ without loss of generality. 
The cyclotron frequency characterizing the separation of Landau levels in the un-tilted model is $\omega_z = B^z/m$, and we define a corresponding energy scale for the in-plane field as $\omega_x = B^x/m$. 
We focus primarily on the regime of strong confinement and small tilt,
\begin{align}
    \omega_x \ll \omega_z \ll \omega_0.
\end{align}

We recover an effective two-dimensional model when electrons are energetically restricted to the lowest level of the confining potential via an appropriate choice of the temperature and chemical potential. As the high energy excited states of the confining potential do not contribute at the level of hydrostatics, our results should generalize to other strong confining potentials. An analysis of the effect of more complicated confining potentials on Hall physics is undertaken by Ref. \cite{NonHarmonicPotentials}.

It is convenient to parametrize the range of applicability of the effective two-dimensional model in terms of the small parameters  
\begin{align}
    k = \omega_x/ \omega_z \qquad \text{and} \qquad l = \omega_z/\omega_0.
\end{align} 
In our analytical calculations that follow we will work to maximum order $\mathcal O(k^2) \times \mathcal O(l^2)$.

We solve for the eigenstates of the tilted field Hamiltonian  by closely following and borrowing notation from Refs.~\cite{offertaler2019viscoelastic,yang2017anisotropic}. For simplicity, in what follows we will work with the single-particle Hamiltonian and suppress the particle indices $i,j$. We will restore them explicitly in expressions that require a sum over particles. 
In the case where $k =0$ ($\mathbf{B} = B^z \mathbf{\hat z}$), the spectrum of the single-particle Hamiltonian is determined by two sets of manifestly gauge invariant ladder operators,
\begin{align}
     a &= \frac{1}{\sqrt{2m\omega_z}}(  \pi_x - i \pi_y),\\
    b &= \frac{1}{\sqrt{2m\omega_0}} (\pi_z - i m \omega_0 z),
\end{align}
where $[a,a^\dag]=[b,b^\dag]= 1$, $[a,b]=[a,b^\dag]=0$. 

These operators fail to diagonalize Eq.~\eqref{eq:ManyBodyHam} for nonzero $k$ because the in-plane magnetic field couples in and out-of-plane degrees of freedom via the Lorentz force, and so the ladder operators no longer commute with each other. 
Instead, we have
\begin{align}
    [a,b] = [a,b^\dag] = - \frac{1}{2}k \sqrt{l}.
\end{align}

We decouple the degrees of freedom by introducing transformed ladder operators
\begin{align}
    \alpha = a + \frac{k \sqrt{l}}{2}(b - b^\dag),
\end{align}
such that canonical commutation relations are satisfied with $[\alpha,\alpha^\dag] = [b,b^\dag] = 1,$ $[\alpha,b]=[\alpha,b^\dag]=0.$
We will denote the linear transformation which takes the ($\frac{\pi_\mu}{\sqrt{m}},\omega_0 \sqrt{m}z$) coordinate operators to the $(\alpha^\dag, \alpha,b^\dag, b)$ ladder operators as $M$. 
In matrix notation, we can write
\begin{align}\label{eq:alphabdefs}
    ( \alpha^\dag, \alpha,b^\dag, b)^T &= M (\frac{\pi_x}{\sqrt{m}},\frac{\pi_y}{\sqrt{m}},\frac{\pi_z}{\sqrt{m}}, \omega_0 \sqrt{m} z)^T,\\
    M &= \frac{1}{\sqrt{\omega_z}}\begin{pmatrix}
        \frac{1}{\sqrt{2}} & \frac{i}{\sqrt{2}} & 0 & i \frac{k l}{\sqrt{2}}\\
        \frac{1}{\sqrt{2}} & -\frac{i}{\sqrt{2}} & 0 & -i \frac{k l}{\sqrt{2}}\\
         0&0&\sqrt{\frac{l}{2}}&i \sqrt{\frac{l}{2}}\\
        0&0&\sqrt{\frac{l}{2}}&-i \sqrt{\frac{l}{2}}\\
    \end{pmatrix}.
\end{align}

There is another set of operators, $\{c^\dag,c\}$, which commute with the ladder operators $\alpha$, $b$, and with the Hamiltonian $H$. 
These define the guiding center coordinates in the complex plane, and are given by
\begin{align}\label{eq:cDef}
   c = \alpha^\dag - i \sqrt{\frac{B^z}{2}}(x + i y) .
\end{align}

The Hamiltonian Eq.~\eqref{eq:ManyBodyHam} rewritten in terms of the $\alpha$ and $b$ ladder operators is still not diagonal. 
We find its eigenvalues and eigenstates via a canonical, unitary transformation (Bogoliubov transform) which naturally defines ladder operators $X$ and $Y$ that satisfy
\begin{align}
    [H,X] &= -\omega_1 X,~~~~ [H,X^\dag]= \omega_1 X^\dag,\\
    [H,Y] &= - \omega_2 Y, ~~~~~[H,Y^\dag ] = \omega_2 Y^\dag,\\
    [X^\dag,Y] &= [X,Y]=0,
\end{align}
where $\omega_1,\omega_2$ are the excitation energies of the Hamiltonian which are perturbatively connected to $\omega_z,\omega_0$ by
\begin{align}
    \label{eq:omega1}
    \omega_1 = \omega_z \left(1 - \frac{k^2l^2}{2}+\mathcal O(k^2l^4)\right),\\ \label{eq:omega2}
    \omega_2 = \omega_0 \left(1 + \frac{k^2l^2}{2}+\mathcal O(k^2l^4)\right),\\
    \omega_1^2 + \omega_2 ^2 = \omega_x^2 + \omega_z^2 + \omega_0^2.
\end{align}

The Bogoliubov transformation which transforms the $\{\alpha,b\}$ basis of ladder operators to the diagonal $\{X,Y\}$ basis is given explicitly as 
\begin{align}\label{eq:Umatrix}
    &U= \frac{1}{2\sqrt{\omega_2^2- \omega_1^2}}\begin{pmatrix}
       U_1^+& U_2^+\\
       U_2^- & U_1 ^-
    \end{pmatrix}\\
    U_{1}^\pm =& \begin{pmatrix}
        (\omega_z + \omega_1) \sqrt{\frac{\omega_2^2 - \omega_z^2}{\omega_z \omega_1}} & \pm(\omega_z- \omega_1) \sqrt{\frac{\omega_2^2 - \omega_z^2}{\omega_z \omega_1}},\\
      \pm(\omega_z- \omega_1) \sqrt{\frac{\omega_2^2 - \omega_z^2}{\omega_z \omega_1}} &  (\omega_z + \omega_1) \sqrt{\frac{\omega_2^2 - \omega_z^2}{\omega_z \omega_1}} 
    \end{pmatrix},\\
     U_2^\pm =& \begin{pmatrix}
        \pm(\omega_2+ \omega_z) \sqrt{\frac{\omega_z^2 - \omega_1^2}{\omega_z \omega_2}} & (\omega_2 - \omega_z) \sqrt{\frac{\omega_z^2 - \omega_1^2}{\omega_z \omega_2}}\\
       (\omega_2 - \omega_z) \sqrt{\frac{\omega_z^2 - \omega_1^2}{\omega_z \omega_2}} &  \pm(\omega_2+ \omega_z) \sqrt{\frac{\omega_z^2 - \omega_1^2}{\omega_z \omega_2}}
    \end{pmatrix}.
\end{align}

Where
\begin{align}\label{eq:UDef}
   (X^\dag, X,Y^\dag,Y)^T  = U( \alpha^\dag,\alpha,b^\dag,  b)^T.
\end{align}

The resulting single-particle Hamiltonian is equivalent to each single-particle term in Eq.~\eqref{eq:ManyBodyHam}, and takes the form
\begin{align}\label{eq:tiltHam}
    H = \frac{\omega_1}{2}( X^\dag X + X X^\dag) + \frac{\omega_2}{2}(Y^\dag Y + Y Y^\dag).
\end{align}

Note that although Eqs.~\eqref{eq:omega1} and \eqref{eq:omega2} are given to lowest order in $k$ and $l$, The Bogoliubov transformation Eq.~\eqref{eq:Umatrix} and Hamiltonian \eqref{eq:tiltHam} implicitly define $\omega_1$ and $\omega_2$ to all orders in $k$ and $l$.

For our purposes, $X$ and its Hermitian conjugate are perturbatively connected to the Landau level operators of the isotropic Hall system. 
Indeed, as $k\xrightarrow[]{}0$, $X\xrightarrow[]{} a$  and $\omega_1 \xrightarrow[]{}\omega_z$.
For this reason, we refer to the $X$ operators and the corresponding degrees of freedom as ``quasi-Landau."
Likewise, $Y$ and $Y^\dag$ are perturbatively connected to the raising and lowering operators of the perpendicular confining potential, and in the tiltless limit $k\rightarrow 0$ we have $Y \xrightarrow[]{}b$ and $\omega_2 \xrightarrow[]{}\omega_0$. 
We refer to these operators and their corresponding degrees of freedom as ``quasi-confining potential."

Interestingly, the lowest order correction to the energies appears at order $k^2l^2.$ This is a feature of the coupling between the confining potential and the Landau  modes which, as we will see in section \ref{sec:Bohr}, presents in the symplectic two-form as $\delta=kl$.
Since the energy must be invariant under the transformation $k \xrightarrow[]{}- k$ due to the rotational invariance of the $k = 0$ Hamiltonian, the direction of the tilted component of the magnetic field cannot affect the energy.
Therefore, $k$ must enter into the expression for the energy at even orders. 
Thus, the smallest order of $\delta$ which is consistent with the symmetry argument above is $\delta^2 = k^2l^2.$ 
Note that $\delta$ is independent of $B^z.$
We will see later that there exist shear stresses in the ground state of the TFQH droplet which are not restricted by this symmetry argument and therefore lead to corrections to the stress tensor at lower perturbative order.

\bigskip

\subsection{Calculation of Stress for the TFQH Droplet}\label{sec:StressCalc}

Here, we use the exact solution of the tilted field model to calculate the expectation of the stress operator in the ground state. 
The stress operator is defined through the momentum density continuity equation
\begin{align}\label{eq:MomCont}
  \partial_t g_\nu &+  \partial_\mu (\tau^{3D})^\mu_{~\nu} = f^\text{ext}_\nu,
\end{align}
where $g_\nu$ is the (kinetic) momentum density operator and $\rho(\textbf{r})$ is the density operator, given as
\begin{align}
    g_\nu(\textbf{r}) &= \frac{1}{2}\sum_i \{\delta^3 (\textbf{x}_i- \textbf{r}), \pi_\nu^i\},\\
    \rho(\textbf{r}) &= \sum_i \delta^3(\textbf{x}_i - \textbf{r}).
\end{align}

The external force density $f^{\text{ext}}$ acts as a source of momentum. 
In our model, $f^{\text{ext}}$ is given by
\begin{align}
f^{\text{ext}}_\nu &=- \frac{1}{m }\epsilon_{\nu\mu\rho}g^\mu(\textbf{r}) B^\rho - m \omega_0^2 z \delta^3_{\nu}\rho(\mathbf{r}),\label{eq:fextdef}
\end{align}
as the sum of the Lorentz force density [first term in Eq.~\eqref{eq:fextdef}] and the force density due to the confining potential [second term in Eq.~\eqref{eq:fextdef}]. 
We may find the time derivative of $g_\nu$ using the Heisenberg equations of motion
\begin{align}
   \partial_t g_\nu = -i [H,g_\nu],
\end{align}
with the Hamiltonian given by Eq.~\eqref{eq:ManyBodyHam}.
Solving Eq.~\eqref{eq:MomCont} for the stress density allows us to define~\cite{offertaler2019viscoelastic}
\begin{align}\label{eq:StressOpp}
    (\tau^{3D})^\mu_{~\nu} = \sum_i^N \frac{\eta^{\mu\rho}}{4m}\left\{ \{ \pi^i_\rho,\delta^3(\textbf{x}_i- \textbf{r})\}, \pi_\nu^i\right\}.
\end{align}
Eq.~\eqref{eq:StressOpp} is the general stress density operator which is independent of $k$ and $l$ apart from minimal coupling. 

Note that Eq.~\eqref{eq:MomCont} does not uniquely define $(\tau^{3D})^\mu_{~\nu}$, since a divergenceless quantity may be added without changing the continuity equation: $(\tau^{3D})^\mu_{~\nu} \xrightarrow[]{}(\tau^{3D})^\mu_{~\nu} + \epsilon^{\mu\rho\lambda}\partial_\rho\Omega_{\nu\lambda}$. 
Nevertheless, when we integrate over space to obtain the extensive stress operator, ambiguous total derivative terms do not contribute and we find
\begin{align}
    T^\mu_{~\nu}= \int d^3r(\tau^{3D})^\mu_{~\nu}(\mathbf{r}) =  \sum_i^N \frac{\eta^{\mu\rho}}{2m}\{\pi^i_\rho,\pi^i_\nu\}.
\end{align}

Recall that since $\eta^{\mu\nu}=\delta^{\mu\nu}$, the extensive stress operator is symmetric under an exchange of indices.
For the strictly two-dimensional Hall droplet ($l\xrightarrow[]{}0$), the expectation value in the ground state with $\nu$ filled Landau levels is given by the expectation value of the same operator, with~\cite{bradlyn2012kubo}
\begin{align}\label{eq:2Dpressure}
\begin{split}
    \langle T^\mu_{~\nu}\rangle_0 ^{2D}&\rightarrow \sum_i^N\frac{\eta^{\mu\rho}\eta_{\rho\nu}B^z}{2m}\left\langle f(a_i,a^\dag_i) +2 a_i^\dag a_i + 1\right\rangle_0\\
    &= \delta^\mu_\nu \frac{N_0\omega_z}{2}\sum_{n}^\nu(2n+1) = \delta^\mu_\nu \frac{N_0\omega_z \nu^2}{2} = \delta_\nu^\mu E_0,
    \end{split}
\end{align}
where $f(a_i,a^\dag_i)$ is a function of the operators with zero expectation in the ground state, and $E_0$ is the ground state energy.
Eq.~\eqref{eq:2Dpressure} is precisely the statement that the pressure of the Hall droplet is equivalent to the energy density and there are no shear stresses which is the typical description of a fluid \cite{Kadanoff1963}.

The ground state stress tensor in the tilted field model can be similarly evaluated using the Bogoliubov transform in Eqs.~\eqref{eq:alphabdefs} and \eqref{eq:UDef} to rewrite the $\pi_\mu$ operators in terms of the $X$ and $Y$ operators and then taking the expectation value as was done in Ref. \cite{offertaler2019viscoelastic}.
We will project the 3$\times 3$ tensor onto the plane of confinement by introducing indices $\alpha,\beta\in \{1,2\}$. Doing so, we find
\begin{align}
    \begin{split}\langle T^\alpha_{~\beta}\rangle_0 &=  \frac{N_0 \nu^2}{2 }\omega_1  \delta_\beta^\alpha   + \frac{N_0\nu}{2} k l^2(\nu\omega_1 - \omega_2)\frac{ (\delta_\beta^x B^\alpha - \delta_\beta^\alpha  B^x)}{B^z}\\
   &~~~~~~~~~~~~~~~~~~~~~~~~~~~~~~~~~~~~~~~~~~~~~~ + \mathcal O (k^2l^3)
    \end{split}\\ \label{eq:Tanisotropy}
    &\equiv \underline T  \delta ^\alpha _{~\beta} + \Delta T(\sigma_z)^\alpha _{~\beta}, 
\end{align}
where $\nu $ is the filling factor of the quasi-Landau levels in the ground state and $\sigma^z$ is the third Pauli matrix. 
We assume the chemical potential $\mu < \omega_2$ such that only the lowest quasi-confining potential level is filled. We see that to quadratic order in $k$ and $l$, the extensive stress tensor picks up a traceless component $\Delta T(\sigma_z)^\alpha _{~\beta}$. 
To find the stress area density $\langle\tau^\alpha_{~\beta}\rangle$ we divide the extensive stress by the area $A$ of the droplet in the two-dimensional plane. 
Note that by dividing by the area instead of the volume, we have effectively found the ground state expectation of Eq.~\eqref{eq:StressOpp} integrated over the $z-$direction, leaving us with an effective two-dimensional average stress tensor.
In terms of $k,l$, and other known quantities
\begin{align}\label{eq:anisostress}
      \langle \tau \rangle_0 = \frac{\langle\rho\rangle_0 \omega_z}{2} \begin{pmatrix}
      \nu \left(1 - \frac{k^2l^2}{2}\right) & 0\\
      0 & \nu \left(1 - \frac{3k^2l^2}{2}\right)  +  k^2l
    \end{pmatrix},
\end{align}
where the particle area density is given by $\langle\rho\rangle_0  = \rho_0 \nu = \frac{N_0\nu}{A} = \frac{B^z\nu}{2\pi}$ for $\mu < \omega_2$.
Heuristically, instead of describing the forces acting on an infinitesimal cube, the stress area density describes the forces acting on the faces of a long parallelepiped stretched along the $z-$axis an amount $z_0\gg \sqrt{\hbar/m \omega_0}$ much greater than the characteristic length of the ground state of the confining potential. 

The non-zero traceless contribution to the stress 
\begin{equation}
\Delta \tau \equiv \frac{\Delta T}{A} =\frac{\omega_z k^2l}{4}\langle \rho\rangle_0( 1- l \nu)
\end{equation}
distinguishes the electron liquid of the TFQH droplet from typical liquids which have isotropic pressure in the steady state by virtue of the free movement of particles which flow until pressure is equilibrated. 
Instead, the fluid of the tilted field system is a \textit{strange liquid} which flows to maintain an  equilibrium where compression along the $y$-axis is met with a greater restoring force than compression along the $x-$axis, violating Pascal's law.
In other words, the TFQH droplet supports hyperbolic shear stress in equilibrium. 

It should be noted that stress anisotropy cannot be easily achieved in a simple system like the 2DEG through other methods. 
For example, band mass anisotropy (or equivalently a non-uniform metric) yields a strictly isotropic $\langle T^\alpha_{~\beta}\rangle_0$ \cite{offertaler2019viscoelastic}. 
Additionally, it was shown in Ref. \cite{Huang2022} that discrete rotational symmetries of a lattice potential is not sufficient to induce anisotropic pressure. 
However, anisotropic pressure has been noted to emerge in fluids in the presence of external sources, such as can arise in relativistic heavy ion collisions where thermalization occurs rapidly in the plane perpendicular to the colliding ion beams~\cite{florkowski2008anisotropic,florkowski2011highly,ryblewski2008general}. 
Additionally, anisotropic pressure has been noted in supersymmetric quantum field theory models of confined systems in the presence of a magnetic field, analogous to our tilted-field system~\cite{critelli2014anisotropic,jain2015shear}. 
We will argue that in both the tilted field system and in these quantum field theory systems, anisotropy in the ground state stress originates from a combination of rotational symmetry breaking and quantization. 
In the next section, we show that the pressure is guaranteed to be isotropic if one uses strictly classical arguments and explore the quantum origin of the anisotropic stress in the tilted field system.

\section{Stress from Phase-Plane Projections}\label{sec:classical}

In this section, we ask and answer the following question: from where is the anisotropic $\Delta T$ [Eq.~\eqref{eq:Tanisotropy}] derived?
The first hypothesis we consider is that the anisotropy in the pressure is a direct result of the breaking of rotational symmetry by the in-plane magnetic field. 
This would indicate that the anisotropy is ``located" in the $\pi_\mu$ operators which contain information about the magnetic field through the minimal coupling of the vector potential.
In Sec.~\ref{sec:Boltzmann} we prove  that this interpretation is incomplete. In Sec.~\ref{sec:Bohr} we explore the quantum origin of the anisotropic stress.

\subsection{Isotropy in the Classical Limit and Temperature Dependence}
\label{sec:Boltzmann}

The ground state average considered in the previous section neglects any contributions from the excited states of the TFQH system. In principle, these contributions can be accessed by thermal fluctuations. This can be implemented by replacing the ground state average with a thermal average, such that the former can be recovered by taking the zero-temperature limit of the latter.

At sufficiently high temperatures, the thermal average of a variable becomes its integral over the classical phase space of the system, weighted by a distribution function $f$. The evolution of this distribution function $f$ is determined by the Boltzmann equation:
\begin{align}
    \partial_tf+\frac{\pi^\mu}{m}\frac{\partial f}{\partial x^\mu}+F^{\text{ext}}_\nu\frac{\partial f}{\partial \pi_\nu}&=0\,.
    \label{eq:Boltzmann}
\end{align}
The term $F^{\text{ext}}_\mu$ is the external force acting on each electron, which is given by 
\begin{align}
    F^{\text{ext}}_\mu&=-\frac{1}{m}\epsilon_{\mu\nu\lambda}\pi^\nu B^\lambda-m\omega_0^2z\delta_\mu^3.
\end{align}

Generally speaking, equilibrium is reached when $f$ becomes stationary (steady state). 
In a non-interacting system, this condition is satisfied when $f$ depends only on the single-electron energy $\epsilon = \frac{1}{2m}(\pi_\mu\pi^\mu + m^2\omega_0^2z^2)$. 
This becomes evident once we express the stationary limit of Eq.~(\ref{eq:Boltzmann}) in terms of the Poisson brackets between the distribution function $f$ and the single-electron energy function (Hamiltonian) $\epsilon$, that is,
\begin{align}
    \frac{\pi^\mu}{m}\frac{\partial f}{\partial x^\mu}+F^{\text{ext}}_\nu\frac{\partial f}{\partial \pi_\nu}=\{\epsilon,f\}_P=0.
\end{align}
Assuming that $f(\epsilon)$, this equation becomes
\begin{align}\label{eq:fPoisson}
    f'(\epsilon)\{\epsilon,\epsilon\}_P=0\,,
\end{align}
which is automatically satisfied given the skew-symmetry of the Poisson brackets.

The distribution function $f$ describes the particle density in phase space, which means that the total number of particles is obtained by integrating \(f\) over the entire phase space. Consequently, the momentum space integral of the distribution function provides the 3D particle density. Similarly, integrating the gauge-invariant momentum $\pi_\mu$ over the phase space, weighted by the distribution function, gives us the total momentum of the system, and integrating \(\pi_\mu f\) over momentum space corresponds to the 3D momentum density of the system.

Confinement in the $z$-direction induces an effective two-dimensional dynamics, concentrating both the particle and momentum densities near the $xy$-plane. Since hydrodynamic behavior is observed at length scales much larger than the confinement length, we can capture the effective 2D dynamics by integrating the 3D densities along the $z$-direction. These integrated quantities represent the thermal average equivalents of the particle and momentum density operators, which will be denoted by $\langle\ldots\rangle_{\text{th}}$, i.e.,
\begin{align}   
\langle\rho\rangle_{\mathrm{th}}&=\int\frac{dz\,d^3 \pi}{(2\pi)^3}  f (\epsilon)\,,
    \\
    \langle g_\mu\rangle_{\mathrm{th}}&=\int\frac{dz\,d^3 \pi}{(2\pi)^3} \pi_\mu  f(\epsilon)\,. \label{eq:momentum-dens}
\end{align}
Note that the momentum density defined in Eq.~(\ref{eq:momentum-dens}) vanishes in the equilibrium. This follows from the fact that $f(\epsilon)$ becomes an even function of $\pi_\mu$. 

Hydrodynamic equations can be derived directly from the Boltzmann equation. For this, we must assume that $f$ is a generic non-equilibrium distribution function and only take the limit when $f$ becomes the equilibrium distribution at the hydrodynamic equations. Thus, integrating the Boltzmann equation~(\ref{eq:Boltzmann}) along the $z-$direction and over momentum space gives rise to the continuity equation
\begin{align}
  m \,\partial_t \langle\rho\rangle_{\mathrm{th}} +\partial_\alpha \langle g^\alpha\rangle_{\mathrm{th}}&=0\,.
\end{align} 
Here, we assumed that the distribution function vanishes at $(\pi_\mu,z)\rightarrow\pm\infty$. Similarly, the dynamical equation for the  momentum density can be obtained from the Boltzmann equation~(\ref{eq:Boltzmann}) by multiplying it by the gauge invariant momentum $\pi_\alpha$ and integrating the resulting expression over the momentum space and along the $z-$direction. In the equilibrium limit, this gives us
\begin{align}
    \partial_t \langle g_\mu\rangle_{\mathrm{th}}+\partial_\alpha\langle\tau^\alpha_{~\mu}\rangle_{\mathrm{th}}&=-\frac{\epsilon_{\mu\nu\lambda}}{m}\langle g^\nu\rangle_{\mathrm{th}} B^\lambda-m\omega_0^2\delta_\mu^3\langle z\rangle_{\mathrm{th}}\,,
\end{align}
where 
\begin{align}
\langle\tau^\mu_{~\nu}\rangle_{\mathrm{th}} &= \frac{1}{m}\eta^{\mu\lambda}\int\frac{dz\,d^3 \pi}{(2\pi)^3}  \pi_\lambda \pi_\nu f(\epsilon)
\label{eq:stress-semiclass}
\end{align}
is the fluid stress tensor~\footnote{To be precise, the stress tensor is often defined as 
\[\langle \tau^\mu_{~\nu}\rangle=\int\frac{dz\,d^3 \pi}{(2\pi)^3}  \frac{\pi^\mu \pi_\nu\, f}{m} -\frac{\,\langle g^\mu\rangle \langle g_\nu\rangle}{m\langle\rho\rangle},\] but both definitions coincide in the equilibrium, which is the focus of this paper.}.

The energy $\epsilon$ is an even function of $\pi_\mu$ for all $\mu$, which forces the off-diagonal components of the stress tensor to vanish in equilibrium. At this point, it is worth pointing out that the preceding discussion can be generalized to generic confining potential $V(z)$. Moreover, since electron energy is isotropic in momentum space, the stress tensor becomes proportional to $\delta^\mu_\nu$, i.e., 
\begin{align} 
    \langle \tau^\mu_{~\nu}\rangle_{\mathrm{th}} &= \frac{2}{3}\delta^\mu_\nu\int \frac{dz\,d^3 \pi}{(2\pi)^3}   (\epsilon-V(z))f(\epsilon)\,. \label{eq:stress-general}
\end{align}

So far we have not specified the equilibrium distribution function, apart from being only a function of the single-electron energy. Therefore, this result is independent of the particle statistics, and holds true for classical particles, bosons and fermions. In particular, for large temperatures, the TFQH system behaves like a non-interacting plasma in a harmonic trap and $f(\epsilon)$ becomes the Maxwell-Boltzmann distribution. The equipartition theorem simplifies the expression~(\ref{eq:stress-general}) in this limit since our Hamiltonian [Eq. \eqref{eq:ManyBodyHam}] only possesses quadratic degrees of freedom~\footnote{The equipartition argument may also be used to show that band mass anisotropy also gives rise to an isotropic pressure in the classical limit.}, that is, 
\begin{align}
    \langle \tau^\mu_{~\nu}\rangle_{\mathrm{th}}\rightarrow k_BT\langle \rho\rangle_{\mathrm{th}}\delta^\mu_\nu\,,
\end{align}
where $T$ is the temperature and $k_B$ is the Boltzmann constant. Associating the diagonal part of the stress tensor with the system linear pressure (force per unit of length), we recover the ideal gas law.

At lower temperatures, electron statistics can be accounted for by setting \(f(\epsilon)\) to the Fermi-Dirac distribution function in expression~(\ref{eq:stress-general}). However, this does not capture the quantum result in the zero-temperature limit, indicating that the validity of such an expression breaks down at very low temperatures. This argument demonstrates that statistical ensembles with continuous phase space degrees of freedom possess isotropic pressure regardless of the confining potential or the direction of the magnetic field.

By the same principle as the Bohr-van Leeuwen theorem~\cite{bohr1970studier,Vanleeuwen}, the magnetic field, irrespective of its orientation, cannot affect the stress tensor since it only appears in Eq.~(\ref{eq:stress-semiclass}) through the definition of the $\pi$ coordinates. Thus, the anisotropy observed in the quantum stress must originate from non-classical dynamics.

Indeed, this is the same reason why it is not possible to describe a Bose gas in a box below the critical temperature solely in terms of the Bose-Einstein distribution function~\cite{landau2013statistical, fetter2012quantum}. When the thermal de Broglie wavelength becomes larger than the box size, the system temperature is unable to smear the discreteness of energy levels, and a macroscopic portion of the system condenses in the lowest energy level. 
Because the Bose-Einstein distribution intrinsically assumes the energy levels form a continuum, it cannot resolve individual energy levels. 
In other words, this ``classical" picture breaks down when the system temperature is smaller than the gap between adjacent energy levels. 
For fermions, this phenomenon is observed in both the de Haas-van Alphen effect and in the Shubnikov-de Haas oscillations in electronic conductivity~\cite{lifshitz1956theory,abrikosov2017fundamentals}.

The discreteness of energy levels can be captured by restricting the integral over momentum space. 
For example, the cyclotron gap limits the momentum values accessible by the quantum states. 
For the sake of argument, let us first focus on the isotropic case, where $B^x=0$, and defer the discussion of the tilted magnetic field case to upcoming sections. 
In this particular scenario, the quantum system decouples, and we find that the operators $\tfrac{\pi_z^2}{2m}+\tfrac{m\omega_0^2 z^2}{2}$ and $\tfrac{\pi_\beta^2}{2m}$ have eigenvalues $\omega_0(\kappa+\tfrac{1}{2})$ and $\omega_z(n+\tfrac{1}{2})$, respectively.

To enforce the discreteness of the energy levels, one must impose that the classical counterparts of such operators can only assume these discrete values. 
This can be done by defining 
\begin{widetext}
\begin{align}
    \langle\tau^\mu_{~\nu}\rangle_{\mathrm{th}} &= \int\frac{dz\,d^3 \pi}{(2\pi)^3}  \frac{\pi^\mu\pi_\nu}{m} f(\epsilon) \sum_{n=0}^\infty\delta\left(n+\tfrac{1}{2}-\tfrac{\pi_\alpha^2}{2m\omega_z}\right)\sum_{\kappa=0}^\infty\delta\left(\kappa+\tfrac{1}{2}-\tfrac{\pi_z^2}{2m\omega_0}-\tfrac{m\omega_0 z^2}{2}\right)\,. \label{eq:stress-Poisson}
\end{align}    
\end{widetext}

This expression recovers the quantum expectation value when $T\rightarrow 0$. 
Using the Poisson summation formula
\begin{align}\label{eq:Poisson}
    \sum_{n\in\mathbb Z}\delta(n-x)=\sum_{q\in\mathbb Z}e^{2\pi iqx}
\end{align}
we can rewrite the expression for the stress tensor~(\ref{eq:stress-Poisson}) in terms of the semiclassical expression~(\ref{eq:stress-semiclass}) with the additional correction due to quantum oscillations
   \begin{align}
    \langle\tau^\mu_{~\nu}\rangle_{\mathrm{th}} =& \sum_{s,q\in \mathbb Z}\frac{(-1)^{s+q}}{m}\int\frac{dz\,d^3 \pi}{(2\pi)^3}\,  \pi^\mu\pi_\nu \,f(\epsilon )\nonumber
    \\
    &\times \exp\left[i\pi\left( \frac{s\pi_\alpha^2}{m\omega_z}+q \frac{\pi_z^2+m^2\omega_0^2z^2}{m\omega_0}\right)\right] .\label{eq:stress-oscillation}
\end{align}     
When $T\gg \omega_0,\omega_z$, the oscillation terms with $s,q\neq 0$ get suppressed and the semiclassical expression~(\ref{eq:stress-semiclass}) is recovered. 

Introducing the discreteness of energy levels keeps the stress tensor diagonal for this particular case. 
However, it disrupts the three-dimensional isotropy of the stress tensor at low temperatures, as $\langle\tau^x_{~x}\rangle_{\mathrm{th}}=\langle\tau^y_{~y}\rangle_{\mathrm{th}}\neq \langle\tau^z_{~z}\rangle_{\mathrm{th}}$. In fact, when the quantum oscillations dominate, we have $\langle\tau^x_{~x}\rangle_{\mathrm{th}}=\langle\tau^y_{~y}\rangle_{\mathrm{th}}\approx  l\langle\tau^z_{~z}\rangle_{\mathrm{th}}$, as we will show in Sec.~\ref{sec:Bohr}.

We will see that discreteness of energy levels is also responsible for the predicted anisotropy in the TFQH droplet. 
In the next section, we will apply the quantum constraints to areas of trajectories in phase space using the Bohr-Sommerfeld quantization scheme when $B^x\neq 0$, and derive a relationship between the Bogoliubov transformation [Eq.~\eqref{eq:UDef}] and a rotation of phase space coordinates which decouples the symplectic two-form.

\subsection{Semiclassical Trajectories}
\label{sec:Bohr}

Because the TFQH Hamiltonian [Eq. \eqref{eq:tiltHam}] is quadratic and decouples as two independent quantum harmonic oscillators, we can semiclassically quantize the system using the Bohr-Sommerfeld approach to obtain exact results~\cite{abrikosov2017fundamentals}.
The goal of this subsection will be to find the closed, classical trajectories in phase space and apply the quantization condition to connect to the quantum dynamics.
The classical action is given by 
\begin{align}\label{eq:Action}
    S = \int \left[ \frac{1}{2}\widetilde \xi_{\widetilde A} \widetilde\omega^{\widetilde A\widetilde B}\dot{\widetilde  \xi}_{\widetilde B} - H(\widetilde \xi_{\widetilde A})\right] dt.
\end{align}

\begin{figure}[t]
    \centering
    \includegraphics[width = 8cm]{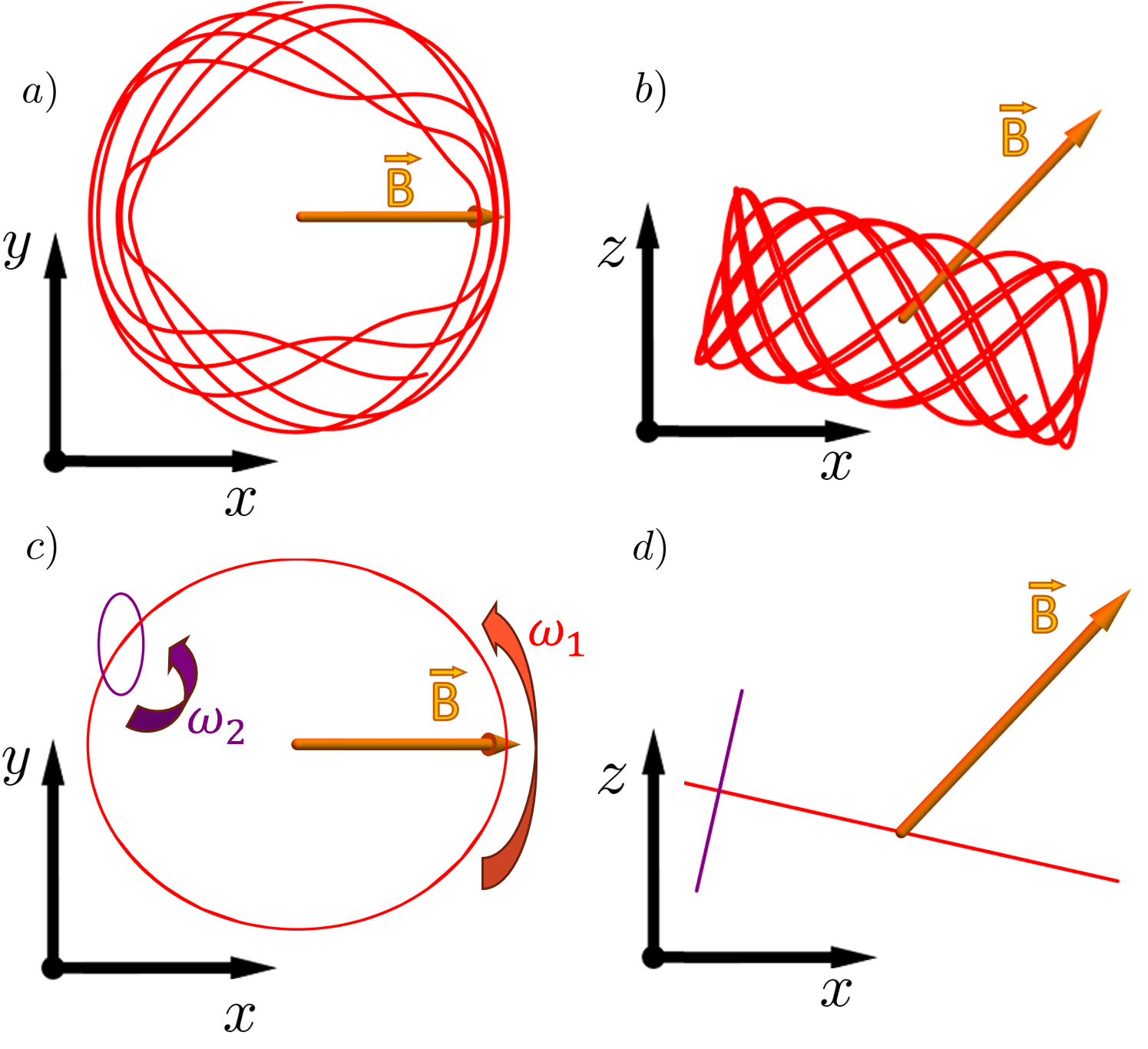}
    \caption{a) and b) A representative trajectory of a classical particle subject to a tilted magnetic field and confining potential along the $z-$axis. $k$ and $l$ are chosen to be 0.95 and 0.5 respectively which are non-perturbative, but exaggerate the effects of the anisotropy. c) and d)  The decoupled orbits of the quasi-Landau (red) and quasi-confining potential (purple) degrees of freedom form ellipses when projected onto the $x-y$ plane which have perpendicular semi-major axes.}
    \label{fig:ClassicalTrajectories}
\end{figure}

\begin{figure}[b]
    \centering
    \includegraphics[width = 9 cm]{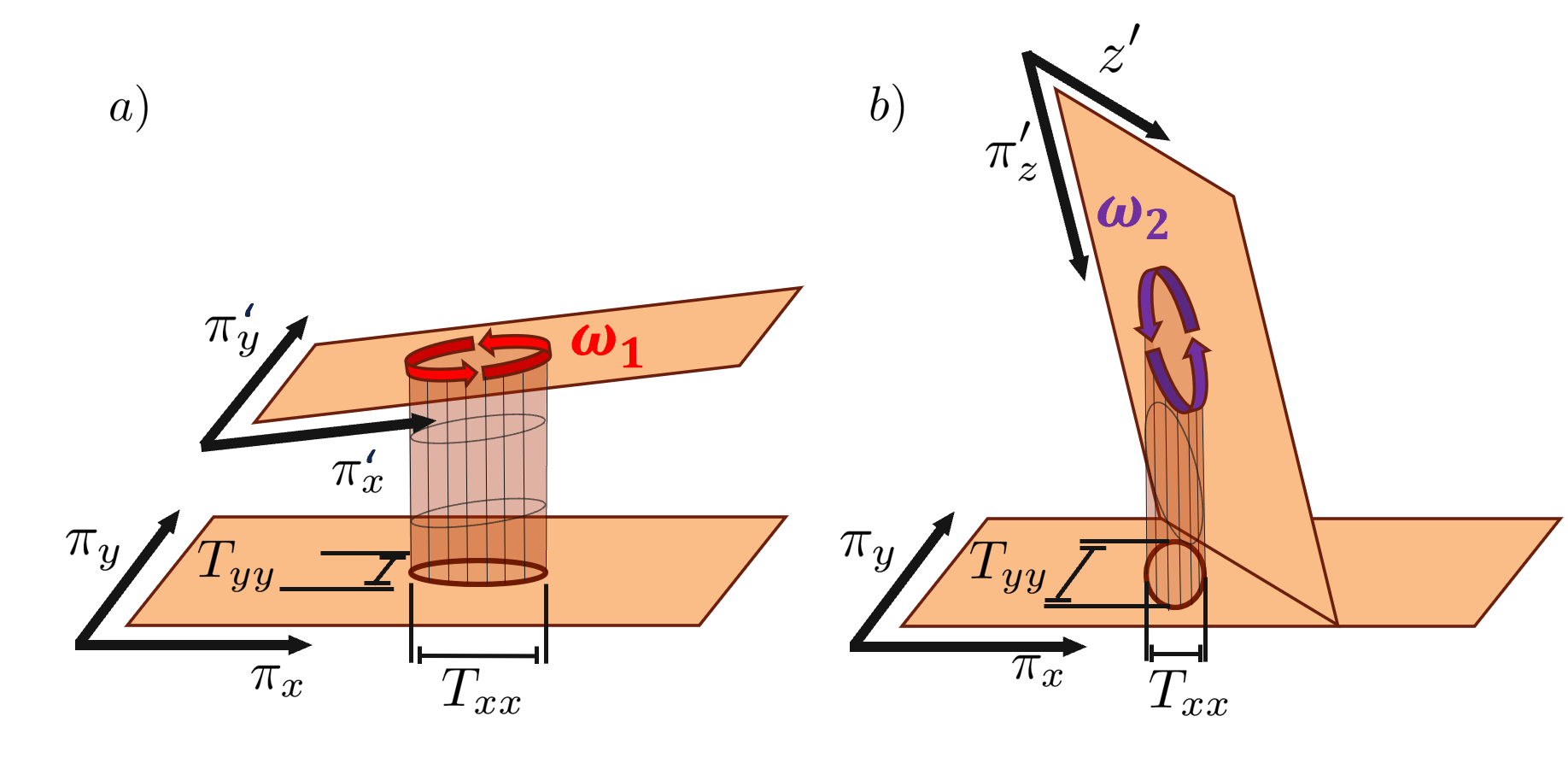}
    \caption{A visual representation of the rotation and projection of the decoupled phase space planes onto the original $\pi_x-\pi_y$ plane. a) For perturbative $k$ and $l$ the rotation is small and the projection of the trajectory is an ellipse with small eccentricity. b) When $k =0$ or $l = 0$ all trajectories on $P_2$ project onto the origin of the plane spanned by $\{\pi_x,\pi_y\}$. 
    In the tilted case, the rotation allows an elliptic projection which, for small 
 values of $k$ and $l$, has minor and major axes proportional to $\{kl^{5/2},kl^{3/2}\}$ respectively.}
    \label{fig:rotate}
\end{figure}

\begin{figure*}
    \centering
    \includegraphics[width = 18cm]{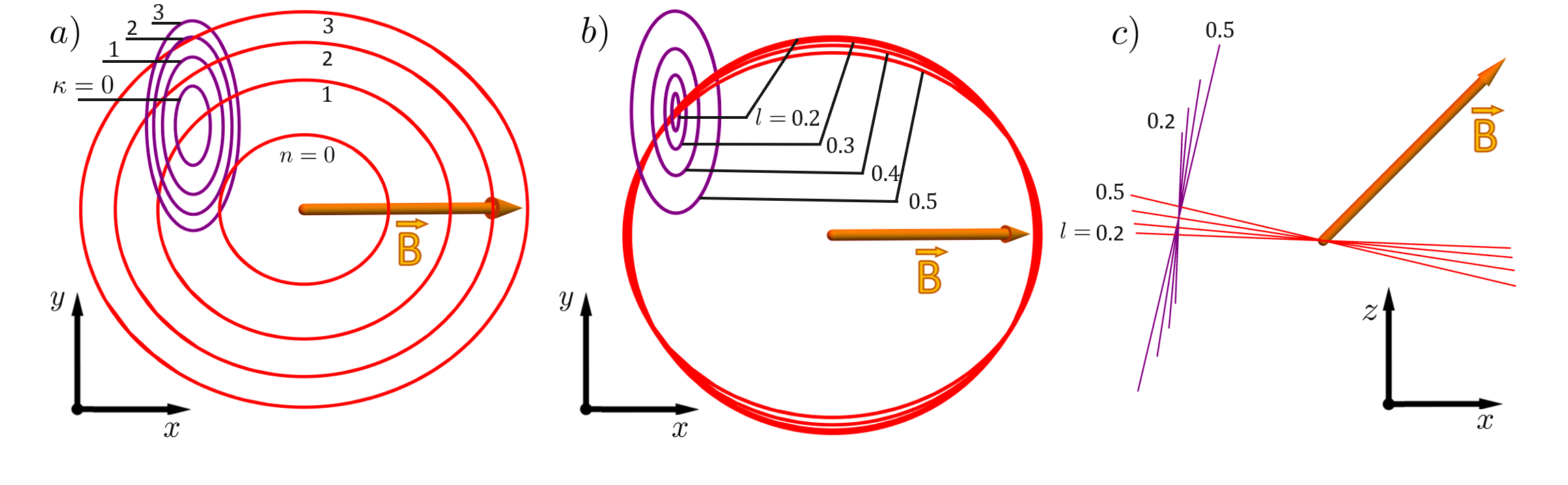}
    \caption{ a) Classical trajectories projected onto the $x-y$ plane where the area swept by the closed path in phase space is quantized via the Bohr-Sommerfeld scheme for given values of $n$ and $\kappa$ in Eq.~\eqref{eq:bohrsommerfeldintegrals}. $k = 1, l = 0.5$ are non-perturbative for demonstrative purposes. 
    b) and c) $x-y$ projection of the ground state ($n = \kappa  = 0$) orbits for $k =1$, $l = \{0.2,0.3,0.4,0.5\}$.}
    \label{fig:TrajectoryDepen}
\end{figure*}

Where $\widetilde A, \widetilde B\in \{1,\dots,6\}$  index phase space coordinates and $\widetilde\omega$ is the symplectic two-form.
We will use coordinates
\begin{align}
   \widetilde  \xi_1 &= \frac{\pi_x}{\sqrt{m}} ,~~~~~~\widetilde\xi_4 = \sqrt{m} \omega_0 x,\\
    \widetilde\xi_2 &= \frac{\pi_y}{\sqrt{m}},~~~~~~\widetilde\xi_5 = \sqrt{m}\omega_0 y,\\
    \widetilde\xi_3 &=\frac{\pi_z}{\sqrt{m}},~~~~~~\widetilde\xi_6=\sqrt{m}\omega_0z,
\end{align}
such that the Hamiltonian is  $H = \frac{1}{2}(\widetilde\xi_1^2 + \widetilde\xi_2^2+\widetilde\xi_3^2+\widetilde\xi_6^2).$  The symplectic form is 
\begin{align}
   \widetilde \omega =& \frac{1}{\omega_0 }\begin{pmatrix}
       0 & \mathbb I\\
        -\mathbb I  & -F/m \omega_0 
    \end{pmatrix}.
\end{align}
and $F$ is the electromagnetic field strength tensor. 
Since Eq.~\eqref{eq:Action} is gauge invariant, the following arguments are necessarily independent of the choice of gauge up to a total derivative inside the action.
Figs. \ref{fig:ClassicalTrajectories}(a) and \ref{fig:ClassicalTrajectories}(b) show a representative trajectory which minimizes the action for fixed values of $k$ and $l$ chosen to be large enough to make the distortion of the orbits visible.

To semiclassically quantize the dynamics we first perform a symplectomorphism
\begin{align}
    \xi_1 &= \widetilde\xi_1~~~~~~ \xi_2 = \widetilde\xi_2\\
    \xi_3 &= \widetilde\xi_3~~~~~~ \xi_4 = \widetilde\xi_6\\
    \xi_6 &= \widetilde\xi_4 + \frac{1}{l}\widetilde\xi_2 +k \widetilde\xi_6\\
     \xi_5 &= \widetilde\xi_5 - \frac{1}{l}\widetilde\xi_1
\end{align}
by defining guiding center variables $\xi_5$ and $\xi_6.$
This linear transformation partially block-diagonalizes the symplectic form, leaving only a weak coupling between coordinate pairs {$(\xi_1,\xi_2)$ and $(\xi_3,\xi_4)$ with magnitude $kl$ (which we referred to as $\delta$ before).
\begin{align}
    \omega &=  \frac{1}{\omega_z}\begin{pmatrix}
        0 & 1 & 0 & k l\\
        -1 & 0 & 0 &0\\
        0 & 0 & 0 & l\\
        -kl & 0 & -l & 0
    \end{pmatrix}\oplus -\frac{\omega_z}{\omega_0^2} i \sigma^y \\
    H &= \frac{1}{2}\xi_{A} \eta^{AB} \xi_B
\end{align}
Where $A,B\in\{1\dots 4\}$ and $\sigma^y$ is the second Pauli matrix.
The guiding center coordinates do not couple to the other coordinates through the symplectic form, nor are they represented in the Hamiltonian. 
We can thus write the action in the $\xi$ coordinates as
\begin{align}
\begin{split}
    S = &\int \left[\frac{\omega_z}{2\omega_0^2}( 
   \xi_5 \dot {\xi_6} -  
   \xi_6 \dot {\xi_5} )\right] dt\\
   + &\int \left[ \frac{1}{2}\xi_A \omega^{AB}\dot{\xi}_B - \frac{1}{2} \xi_A \eta^{AB}
   \xi_B\right] dt.
   \end{split}
\end{align}

The equations of motion require $\dot{\xi}_{5,6}$ to be constant, and since $\xi_{5,6}$ are not present in the Hamiltonian, they do not contribute to dynamics.
Therefore, we will only consider the action of the $\xi_A$ coordinates going forward.

A linear transformation further decouples the symplectic two-form $\omega$ as
\begin{align}\label{eq:diagonalized}
  \xi_A \omega^{AB} \xi_B &=  \xi_A(V^{-1}V)^A_B\omega^{BC}(V^{-1}V)_C^D  \xi_D\\
  &=\xi'_A (\omega')^{AB}\xi'_B
\end{align}
with
\begin{align}\label{eq:Vdef}
    V = Q |\Lambda|^{-1/2}UM,
\end{align}
where $U$ is the Bogoliubov transform defined in Eq.~\eqref{eq:Umatrix}, $M$ is the linear transformation defined in Eq.~\eqref{eq:alphabdefs}, and 
\begin{align}
 Q &=\frac{1}{\sqrt{2}} \begin{pmatrix}
      1&1&0&0\\
      -i &i &0&0\\
      0&0&1&1\\
      0&0&-i&i
    \end{pmatrix}
\end{align}
converts from complex coordinates to their real and imaginary parts.
The matrix $|\Lambda|^{-1/2}= \text{diag}(\sqrt{\omega_1},\sqrt{\omega_1},\sqrt{\omega_2},\sqrt{\omega_2})$ is present to rescale the classical coordinates $\xi'_A$ such that when $k =0, ~\xi'_A = \xi_A.$ 
This shows that decoupling the symplectic form in the classical action is functionally equivalent to the Bogoliubov transform of the quantum Hamiltonian. 
The decoupled symplectic two-form is

\begin{align}
    \omega' = \begin{pmatrix}
        0 &  \frac{1}{\omega_1} & 0 & 0\\
        -\frac{1}{\omega_1}&0& 0 & 0\\
        0 &0&0&\frac{1}{\omega_2}\\
       0&0&-\frac{1}{\omega_2}&0
    \end{pmatrix}.
\end{align}

The frequencies $\omega_{1,2}$ are defined in equations \eqref{eq:omega1} and \eqref{eq:omega2}.
We emphasize that $V\in SO(4)$ should be understood as a rotation in the four-dimensional phase space (neglecting the two dimensions of the guiding center coordinates).
A visual representation of this rotation is presented in Fig.~\ref{fig:rotate}.

The decoupled coordinates are related to the quantum solution by 
\begin{align}
    &\xi'_1 = \sqrt{2\omega_1} \,\text{Re}[X_c],~~~~ \xi'_2 =\sqrt{2\omega_1}\, \text{Im}[X_c],~~~~\\
    &\xi'_3 =\sqrt{2\omega_2} \,\text{Re}[Y_c],~~~~~ \xi'_4 =\sqrt{2\omega_2} \,\text{Im}[Y_c].
\end{align}

The coordinates $X_c$ and $Y_c$ are the classical analogues of the quantum operators defined in Eq. \eqref{eq:UDef} by the Bogoliubov transformation.
Again, the rescaling by $\sqrt{\omega_{1,2}}$ is necessary because the canonical commutators $[X,X^\dag]=[Y,Y^\dag] =1$ while the Poisson brackets of $\xi'_A$, defined through 
\begin{align}
    \omega' = \frac{1}{\omega_1}d\xi'_1 \wedge d\xi'_2 + \frac{1}{\omega_2}d\xi'_3\wedge d\xi'_4,
\end{align}
are 
\begin{align}
    \{\xi'_1,\xi'_2\}_P = \omega_1,~~~~~\{\xi'_3,\xi'_4\}_P = \omega_2
\end{align}

The decoupled coordinates define linearly independent phase space planes spanned by $\{\xi'_1,\xi'_2\}(\{\xi'_3,\xi'_4\} )$ and denoted $P_1(P_2).$
The decoupling of the modes can be seen in Figs. \ref{fig:ClassicalTrajectories}(c) and \ref{fig:TrajectoryDepen} where the projection of the (slow) orbit of the quasi-Landau degree of freedom onto the $x-y$ plane has morphed from a circle in the $kl =0$ limit to an ellipse.
Likewise, the (fast) orbit of the quasi-confining potential mode, which in the $k=0$ limit projects to a single point (the origin is the only point where $P_1$ and  $P_2$ intersect), now projects to a small ellipse.

The Hamiltonian, being only a function of the magnitude of the vector $\xi_A$, has an $SO(4)$ rotational symmetry. 
Therefore, all rotations in the four-dimensional phase space leave the Hamiltonian invariant, and we have
\begin{align}
    H = \frac{1}{2}\xi'_A \eta^{AB}\xi'_B.
\end{align}
The equations of motion resolve into those of two decoupled harmonic oscillators.
\begin{align}
    \dot \xi_1' &= -\omega_1 \frac{\partial H}{\partial \xi'_2} = - \omega_1\xi'_2~~~~~~~
     \dot \xi_2' = \omega_1 \frac{\partial H}{\partial \xi'_1} = \omega_1 \xi'_1 \label{eq:eom1}\\
      \dot \xi_3' &= -\omega_2 \frac{\partial H}{\partial \xi'_4} = - \omega_2 \xi'_4~~~~~~~
       \dot \xi_4' = \omega_2 \frac{\partial H}{\partial \xi'_3} = \omega_2 \xi'_3\label{eq:eom2}
\end{align}

An explicit formulation of the classical velocities in terms of the initial phase space coordinates is given in Appendix \ref{app:ClassicalEOM}.

Bohr-Sommerfeld quantization places a constraint on the phase space area enclosed by the trajectories that solve Eqs.~\eqref{eq:eom1} and \eqref{eq:eom2}; Following Ref.~\cite{abrikosov2017fundamentals} we have the quantization conditions
\begin{align}\label{eq:bohrsommerfeldintegrals}
    \oint \xi'_2 d\xi'_1 &= \omega_1 (n + \tfrac{1}{2}),~~~~\oint\xi_4' d\xi'_3 = \omega_2 (\kappa + \tfrac{1}{2}),
 \end{align}
where $n,\kappa\in \mathbb Z^\geq$ are non-negative integers. 
We can reformulate the quantization condition in terms of a time average using the equations of motion to find
\begin{align}
     \int_0^{1/\omega_1} \xi'_2 \xi_2' \omega_1 dt &= \omega_1 (n+ \tfrac{1}{2}),\\
     \int_0^{1/\omega_2}\xi_4' \xi'_4 \omega_2dt &= \omega_2 (\kappa + \tfrac{1}{2}).
\end{align}

Under the ergodic assumption, a particle will encounter every state in the single-particle phase space in the large time limit with probability determined by the weight $f$~\cite{kardar2007statistical}.
This means that for any variable $h$ we may replace
\begin{align}
    \int \left(\frac{d^3x d^3 \pi }{(2\pi)^3} f\right)  h = \lim_{t\xrightarrow[]{}\infty}\frac{1}{t}\int_0^tdt' h(t'),
\end{align}
where $h(t)$ is the trajectory for $h$ which minimizes the action.
From equation \eqref{eq:stress-semiclass}, the single particle stress is given by
\begin{align}\label{eq:classicalstress}
   \overline  T^\mu_{~\nu} = \lim_{t\xrightarrow[]{}\infty} \frac{1}{tm}\int_0^t dt' \pi^\mu(t')\pi_\nu(t').
\end{align}
Finally, using Eq.~\eqref{eq:classicalstress} and the transformation $V$ [defined in Eq.~\eqref{eq:Vdef}] from the primed to unprimed variables allows us to derive the contribution to the semiclassical time-averaged stress tensor for a single particle.
\begin{align}\label{eq:BSStress}
    \overline T^\mu_{~\nu} &= \eta^{\mu\eta}S_\eta^A\left(\omega_1 (n+ \tfrac{1}{2})(\eta_1)^{A}_B + \omega_2 (\kappa + \tfrac{1}{2})(\eta_2)^A_B\right) S^B_\nu,
\end{align}
where $\eta_{1}$ and $\eta_2$  project onto the phase space planes $P_1$ and $P_2$, respectively, and are given to order $\mathcal O(k^2)\times \mathcal O(l^2)$ as
\begin{align} \label{eq:project1}
   \eta_1&\approx \begin{pmatrix}
    1 & 0 & kl^2 & 0\\
    0 & 1 - k^2l^2 & 0 & k l\\
    k l^2 & 0 & \mathcal{O}(k^2l^4) & 0\\
    0 & k l & 0 & k^2 l^2
    \end{pmatrix},\\\label{eq:project2}
     \eta_2&\approx \begin{pmatrix}
    \mathcal{O}(k^2l^4) & 0 & -kl^2 & 0\\
    0 &  k^2l^2 & 0 & -k l\\
    -k l^2 & 0 & 1 & 0\\
    0 & -k l & 0 & 1 
 - k^2 l^2
    \end{pmatrix}.
\end{align}

The matrix $S$ projects from the four-dimensional phase space onto three-dimensional momentum space, and is given explicitly as
\begin{align}\label{eq:Sdef}
    S = \begin{pmatrix}
        1 &0&0\\
        0 & 1 &0\\
        0 & 0 & 1\\
        0 & 0 & 0
    \end{pmatrix}.
\end{align}

As projection operators onto orthogonal phase space planes, $\eta_1$ and $\eta_2$ satisfy
\begin{align}
    \eta_1 + \eta_2 = \mathbb I,~~~~~\eta_1\eta_2 = 0.
\end{align}

$\eta_2$, which projects onto the $\pi_z-z$ phase space plane when $k = 0$, now projects onto the rotated $P_2$ plane which has some overlap with the $\pi_x-\pi_y$ phase space plane. 
Projecting a circle of unit area from $P_2$ to the $\pi_x-\pi_y$ plane results in an ellipse of area proportional to $k^2l^4$.
This can be seen diagramatically in figure \ref{fig:rotate} and for representative trajectories in figure \ref{fig:TrajectoryDepen}(a).

The average stress area density is given by summing Eq.~\eqref{eq:BSStress} over $N$ particles and dividing by the area, yielding
\begin{align}\label{eq:3Dtau}
    \overline  \tau
     &=\frac{\rho_0 \nu}{2} \begin{pmatrix}
      \nu\omega_1  & 0 & k l^2\Delta, \\
      0 & \nu \omega_1 - k^2l^2\Delta   &0\\
       k l^2 \Delta  & 0 &\omega_2
    \end{pmatrix},\\
    \Delta &= \nu \omega_1 - \omega_2.
\end{align}

By projecting onto the plane of confinement by restricting the indices of $\overline \tau$ to $\alpha,\beta \in\{1,2\}$, we recover Eq.~\eqref{eq:anisostress} .
In equation \eqref{eq:3Dtau}, the chemical potential is restricted to $\mu< \omega_2$.
For general filling (i.e. for general values of $\mu$) we must sum over the states using the Fermi distribution function at zero temperature.
\begin{align}
    \overline\tau &= \frac{\rho_0}{2}\sum_{n,\kappa=0}\Theta(\mu- n\omega_1 - \kappa\omega_2)\nonumber \\
    &\times  \begin{pmatrix}
      (2n+ 1)\omega_1  & 0 & k l^2\Delta \\
      0 & (2n+ 1) \omega_1 - k^2l^2\Delta   &0\\
       k l^2 \Delta  & 0 &(2\kappa+ 1)\omega_2
    \end{pmatrix}\label{eq:semiclassicalstressgeneralmu}\\ 
    \Delta &= ( 2n+ 1)\omega_1 - (2\kappa +1)\omega_2\label{eq:DeltaDef}
\end{align}

Note that {Eq.~\eqref{eq:semiclassicalstressgeneralmu} reveals the presence of a ground state $x-z$ stress proportional to the same parameter $\Delta$ of the in-plane stress anisotropy.

Furthermore, the three-dimensional stress tensor reveals that the ratio between the in-plane pressure and the out-of-plane pressure for the quantum system is approximately $1/l$. This is a result of the same quantum effect described above, however, the out-of-plane stress is not a physically measurable quantity due to the presence of the confining potential in the momentum continuity equation (Eq.~\eqref{eq:MomCont}) which acts as a source for $g_z.$ Since, confinement necessitates the total flow in the $z-$direction be zero, the out-of-plane stress serves to balance the external, confining force in the momentum continuity equation in the steady state.

Of course, tilting the magnetic field is not necessary for the anisotropy in the in and out-of-plane stress to appear because $SO(3)$ symmetry is broken by the confining potential, as noted in our discussion at the end of Sec.~\ref{sec:Boltzmann}.
In the tiltless case, the stress density in the plane is $\langle\tau^{x}_{~x}\rangle_0=\langle\tau^{y}_{~y}\rangle_0=\frac{1}{2}\rho_0\nu^2\omega_z$ while the out of plane stress density is $\langle\tau^{z}_{~z}\rangle=\frac{1}{2}\rho_0\nu\omega_0.$
The ratio of $\langle\tau^{x}_{~x}\rangle_0/\langle\tau^{z}_{~z}\rangle_0$ is equal to $\nu l$ for $\mu < \omega_0.$
For $\omega_z\ll \omega_0$, as $\mu\xrightarrow[]{}\omega_0$ then $\nu\xrightarrow[]{}1/l$ and approximate isotropy is recovered. 

\bigskip

\subsection{Interpretation}
\label{sec:EffMet}

We may now answer the question as to the origin of the anisotropy in the stress tensor for the TFQH system: quantum anisotropic pressure is a direct result of the quantization of phase space trajectories. 
It is, after all, the necessity that $n,\kappa\in \mathbb Z^\geq$ which implies that $\Delta \tau\neq 0$ for incommensurate $\omega_1$ and $\omega_2.$ 

If we allow the spacing of the energy levels to become infinitesimal, by considering  a continuous spectrum or working in the classical limit $\hbar \xrightarrow[]{}0$, then $\kappa, n\xrightarrow[]{}\infty$ such that  $\lim_{\omega_{1,2}\xrightarrow[]{}0}\omega_1(n + \tfrac{1}{2}) + \omega_2 (\kappa + \tfrac{1}{2}) = \mu$ for states on the Fermi surface.
The average over states becomes an integral,
\begin{align}
    \langle \Delta\rangle = \int_0^\infty\frac{dx dy}{\omega_1\omega_2} \Theta(\mu - x - y) 2(x - y)=0
\end{align}

Where we used $\lim_{\hbar \xrightarrow[]{}0}n \omega_1 = x,~\lim_{\hbar \xrightarrow[]{}0}\kappa \omega_2 = y,$ and the integral is zero by symmetry.

\begin{figure}[b]
\includegraphics[width=\columnwidth]{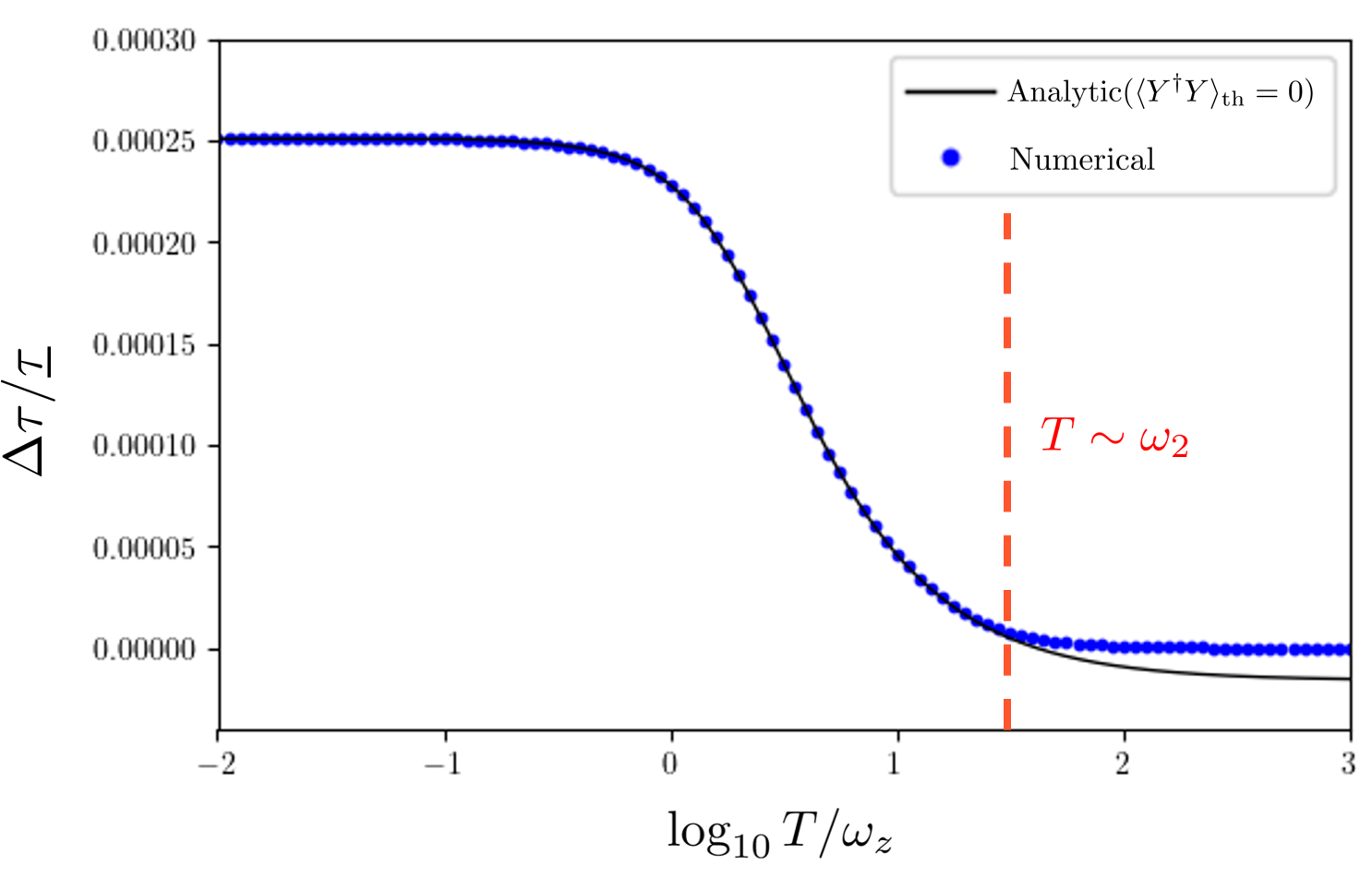}
\caption{The dimensionless measure of anisotropy as a function of temperature $\langle\Delta \tau\rangle/\langle\underline \tau\rangle \equiv \frac{\langle  \tau^{y}_{~y}- \tau^{x}_{~x}\rangle_{\mathrm{th}}}{\langle \tau^{x}_{~x}+ \tau^{y}_{~y}\rangle_{\mathrm{th}}}$. 
An approximate analytic solution constraining $\langle Y^\dag Y\rangle_{\mathrm{th}} =0$ (black) closely fits the numerical Fermi-Dirac sum (blue) for temperatures $T\ll \omega_2$. 
The red dashed line shows the regime where $T$ becomes on the order of $\omega_2$, and thus $\langle Y^\dag Y \rangle_{\mathrm{th}}\neq 0.$ The calculation of $\langle \tau^{\mu}_{~\nu}\rangle_{\mathrm{th}}$ under the $\langle Y^\dag Y \rangle_{\mathrm{th}}=0$ approximation is given in appendix \ref{app:Analytic}. 
In units of $\omega_z$ the parameters used are $\omega_1 \approx 1, \omega_2 \approx 100, \mu= 5.5, k = 0.4,l = 0.01$.}\label{fig:TempDepen}
\end{figure}

An analysis of the stress, thermally weighted via the Fermi-Dirac distribution at non-zero temperature confirms the return of isotropy in the classical (large temperature) limit. 
One important question is: what temperature is large enough to recover the classical isotropic stress?
We plot a dimensionless measure of the anisotropy 
\begin{equation}
\Delta \tau/\underline \tau = \frac{\langle  \tau^{y}_{~y}- \tau^{x}_{~x}\rangle_{\mathrm{th}}}{\langle \tau^{y}_{~y}+  \tau^{x}_{~x}\rangle_{\mathrm{th}}}
\end{equation}
in Fig.~\ref{fig:TempDepen}.
The crossover regime occurs when $T\gg \omega_1$ which is approximately independent of $l$ for $l < 1$ and $\mu < \omega_2$.
An analytic solution is calculated in Appendix \ref{app:Analytic} using the approximation $\langle Y^\dagger Y\rangle_{\mathrm{th}} =0$.

Although the analytic solution is an approximation, it demonstrates that the crossover to isotropy, which begins around $T \approx 2 \omega_z$, is independent of the confining potential states since the approximation by construction does not consider contributions from excited quasi-confining potential states. 
Under this approximation, any crossover strictly originates from the thermalization of the quasi-Landau degrees of freedom. 
At large temperatures, where $T\approx \omega_2$ and $\langle Y^\dag Y \rangle_{\mathrm{th}}\neq 0$, we expect disagreement between the numerical and analytical results which is seen in the figure around $T \approx 20\, \omega_z$. 
The analytical result becomes negative in the large temperature limit, but this is an unphysical artifact of the $\langle Y^\dag Y\rangle_{\mathrm{th}}=0$ approximation, which breaks down for $T\approx\omega_2$. 
As seen in the numerical calculation} the filling of excited quasi-confining potential modes at high temperatures conspires to ensure that the dimensionless anisotropy is positive for all temperatures, $l< 1,$ and $\mu < \omega_2.$

Furthermore, because the difference of the in-plane and out-of-plane stress is also proportional to $\Delta$, the ratio 
\begin{align}
   \lim_{T\xrightarrow[]{}\infty} \left\langle\frac{\tau^x_{~x}+\tau^y_{~y} - 2\tau^z_{~z}}{\tau^x_{~x}+\tau^y_{~y} + 2\tau^z_{~z}} \right\rangle_{\text{th}}= 0
\end{align}
also approaches zero in the classical limit.

Let us underline the strangeness of these conclusions.
{The lack of rotational symmetry in our TFQH model is a necessary but not sufficient condition for the emergence of stress anisotropy.}
The equations of motion are such that the quasi-Landau mode and the quasi-confining potential mode compensate each other, and the classical pressure {at high temperature} is isotropic despite the broken symmetry. 
It is only when the energy spectra of these modes are {discrete due to quantum effects} that the ability of one harmonic degree of freedom to compensate the stress of another is hindered. 

For physical intuition, let us return to Fig.~\ref{fig:TrajectoryDepen}. 
Note that the slow trajectory (in red)  traces out an ellipse oriented along the $x-$axis while the fast trajectory (in purple) traces out an ellipse oriented along the $y-$axis. 
An elliptic trajectory contributes unequally to the single-particle stress by biasing its contribution in the direction of the semi-major axis, however, when averaged over initial positions in phase space using f, the biases of the two trajectories are canceled to yield an isotropic average stress tensor.
In the quantum ground state, however, the area of the fast trajectory is bounded from below. 
Therefore, the fast motion yields a residual anisotropy in the $yy$ component of the stress.

This effect of quantum shear stresses resembles Landau diamagnetism\cite{LandauDiamagnetism,Kahn1970} which gives a quantum correction to the classical Bohr-van Leeuwen theorem. 
Recall that the Bohr-van Leeuwen theorem shows that the classical distribution function of an ensemble of spinless charged particles is not a function of the magnetic field, and therefore the magnetization and magnetic susceptibility must vanish.
The proof relies on the fact that the magnetic field, which has no associated scalar potential, only enters into the classical Hamiltonian via the relation between canonical and kinetic momentum, and so can be eliminated from classical statistical averages via a redefinition of the phase space integration variables. 
Landau diamagnetism arises as a quantum correction derived from the quantization condition and the non-commutativity between kinetic momentum operators $\pi_\mu$. 
For a two-dimensional electron gas, the Fermi-Dirac distribution picks up a semiclassical constraint term due to the quantization conditions,
\begin{align}
\begin{split}
    f(\epsilon) &\xrightarrow[]{}\\
    \sum_{n}&\delta\left( (n + \tfrac{1}{2}) - \frac{1}{2B^z\hbar}(\pi_x^2+\pi_y^2)\right) f(\epsilon).
    \end{split}
\end{align}

Because the constraint term depends on the magnetic field, the quantum gas has a nonvanishing magnetic susceptibility which is forbidden in the classical gas.
The inherently quantum nature of Landau diamagnetism is made manifest due to its proportionality to $\hbar$. 
The magnetization at zero temperature is~\cite{Kahn1970}
\begin{align}
    M \equiv -  \frac{\partial F}{\partial B^z}= -\frac{\nu \hbar}{2m},
\end{align}
where $F$ is the free energy.
Similarly, the anisotropic stress $\Delta$ from Eq.~\eqref{eq:DeltaDef} vanishes in the classical limit.
\begin{align}
    \lim_{\hbar \xrightarrow[]{}0} \langle\Delta\rangle_{0} = 0
\end{align}

The resemblance of Landau diamagnetism and our argument for anisotropic pressure is not coincidental since the pressure of the strictly two-dimensional Hall droplet is a function of the magnetization \cite{bradlyn2015lowenergy} via
\begin{align}
  \langle  T ^\alpha_{~\beta} \rangle_0= \delta^\alpha_\beta \left( N B^z M_z -E_0   \right).
\end{align}

Although Landau diamagnetism and the anisotropic quantum stress are related, we underline that the latter does not require the former and may arise in any system where rotational symmetry is broken and energy levels are discrete.
Indeed, a magnetic field is not necessary, however the macroscopic degeneracy of the quantum Hall droplet enhances the effect.

Our arguments suggest that it is strictly necessary for at least one of the energy modes to be gapped, but not that the chemical potential lie in a gap. 
For example, anisotropic stress persists if the quasi-Landau levels are broadened such that the ground state is a partially-filled quasi-Landau level.

Finally, we note that we may reinterpret our solutions to the equations of motion in Sec.~\ref{sec:Bohr} in terms of an effective geometry.
Decoupling the Hamiltonian \eqref{eq:tiltHam} into linear combinations of independent coordinates using equations \eqref{eq:project1} and \eqref{eq:project2} yields
\begin{align}\label{eq:PhasePlaneHam}
    H =\frac{\eta_1^{AB}}{2m}\pi_A\pi_B + \frac{\eta_2^{AB}}{2m}\pi_A\pi_B,
\end{align}
where $\pi_4 = m\omega_0 z.$
The projection operators $\eta_1$ and $\eta_2$ enter as effective (induced) metrics on the phase-space planes $P_1$ and $P_2$ introduced in Sec.~\ref{sec:Bohr}.
Therefore, it is natural to suspect a two-dimensional effective theory with an effective metric might reproduce the linear response of the TFQH droplet.
Attempts to this end have been tried using the three degrees of freedom of an effective two-dimensional metric, mass rescaling, and an \textit{anisotropic metric}, which couples to the anisotropy in the action through the frame fields~\cite{offertaler2019viscoelastic}.
However, none are completely successful at yielding a theory which is simultaneously $U(1)$ gauge symmetric in the electromagnetic field, contains a stress tensor which is minimally coupled  to the frame fields, and reproduces the correct ground state current, viscosity, and hydrostatic stress.
Eq.~\eqref{eq:PhasePlaneHam} shows that two effective metrics are required to fully capture the dynamics of the TFQH system.

\bigskip

\section{Discussion}

In this work, we have re-examined the anomalous anisotropic stress found in the simplest toy model of the TFQH droplet. 
We showed that, remarkably, the breaking of rotational symmetry by the in-plane magnetic field is necessary but insufficient to generate anisotropy in the ground state average stress tensor, as all classical theories of the same form as Eq.~\eqref{eq:ManyBodyHam} in the thermodynamic limit possess a strictly isotropic hydrostatic stress tensor.
Rather, anisotropic pressure is a purely quantum effect.
Because of the simplicity of the model, we were able to explain this effect in terms of semiclassical arguments: quantization of classical trajectories and projections in phase space. 
Crucially, while our solution makes use of the exact solvability of the  model, the qualitative features of our results are not dependent on the simplifications made.

We believe a possible experimental avenue for indirectly confirming these effects could be to measure the zero bias conductance of a 2DEG connected to a superconducting lead in the presence of a magnetic field which is strong enough to cause Landau quantization, yet not strong enough to quench the superconducting state.
Observations have shown a peak in the conduction under zero bias voltage\cite{Moore1999,Wang2021Andreev} which is attributed to Andreev scattering at the interface induced by the edge state\cite{Takagaki1998Transport,Takayanago1998Andreev}.
The edge state can be understood in terms of classical trajectories which, as we have seen, is modified in the presence of an in-plane magnetic field. 
Therefore, it is reasonable to assume the rate of Andreev scattering might also be predictably modified.

Additionally, our formulation of the hydrostatic stress in terms of projections onto phase space planes is particularly well suited for studying the effect of non-trivial Berry curvature on the stress, which is another possible avenue for future work. 

Going forward, it is natural to examine hydrodynamics of the same TFQH model in an attempt to characterize the anisotropic response in the presence of the tilted field. 
In particular, both the density and the flow profile along the confining direction can contribute to the effective two-dimensional hydrodynamic equations of the quantum Hall system. 
Averaging over the confining direction can, in principle, generate additional terms in the effective in-plane dynamics. 
This is evident when confining a viscous fluid within a thin layer between two parallel plates, also known as the Hele-Shaw setup~\cite{reynolds2022hele, smith1969investigation, miles2019active}. 
We defer a full examination of Hele-Shaw flow in a tilted magnetic field to subsequent work.

\bigskip

\begin{acknowledgments}

The authors thank Xiaoyang Huang and Andrew Lucas for a helpful discussion. 
The work of I.~O. and B.~B. was supported by the Alfred P.~Sloan foundation, and the National Science Foundation under grant DMR-1945058.
\end{acknowledgments}

\begin{appendix}

\bigskip

\section{Classical Equations of Motion}
\label{app:ClassicalEOM}

As discussed in Sec.~\ref{sec:classical}, classical theories of the TFQH droplet do not possess anisotropic pressure. 
The classical trajectories calculated in \ref{sec:Bohr} can be found via solving the equations of motion for different choices of the initial velocities $mv_{\mu0}=\pi_\mu(t=0)$ and the initial value of $z(0)=z_0$. The explicit form of the trajectories are:
\begin{widetext}

\begin{align}
    \pi_{x1}(t) &= v_{x0 } \cos(\omega_1 t)- k z_0 \omega_z (1 + l^2)\sin(\omega_1 t)  + v_{z0} kl^2 \cos(\omega_1 t) + v_{y0}(1 - \frac{k^2l^2}{2})\sin(\omega_1 t), \\
    \pi_{y1}(t) &= v_{y0}(1 - k^2l^2 )\cos(\omega_1 t)- v_{x0}(1 - \frac{k^2l^2}{2})\sin(\omega_1 t) - v_{z0} kl^2\sin(\omega_1 t)- k\omega_z z_0 (1 +  l^2)\cos(\omega_1 t), \\
    \pi_{z1}(t)&= v_{y0}kl^2\sin(\omega_1 t) + v_{x0}kl^2\cos(\omega_1 t) - z_0 \omega_z k^2l^2 \sin(\omega_1 t), \\
    \pi_{x2} (t) &= - v_{z0}kl^2 \cos(\omega_2 t)+ z_{0}\omega_z k l \sin(\omega_2 t), \\
    \pi_{y2}(t) &= v_{z0}kl \sin(\omega_2 t) + v_{y0}k^2l^2 \cos(\omega_2 t)  + kz_0 \omega_z (1 +  l^2)\cos(\omega_2 t), \\
    \pi_{z2 }(t) &= - v_{y0}kl \sin(\omega_2 t) + v_{z0}\cos(\omega_2 t) - v_{x0} kl^2 \cos(\omega_2 t) - z_0 \omega_z /l (1 - k^2l^2/2)\sin(\omega_2 t),
\end{align}
\end{widetext}
to order $\mathcal O(k^2)\times \mathcal O(l^2).$ $\{v_{x0},v_{y0},v_{z0},z_0\}$ are initial values, and the total velocity is the sum of the two degrees of freedom with periodicity $\omega_1$ and $\omega_2$ (e.g. $m v_x(t) = \pi_{x1}(t) + \pi_{x2}(t)$).
As we have seen in Eq.~\eqref{eq:classicalstress}, the classical single-particle stress is a time average. 
Thus, there are no cross terms as long as $\omega_1\neq \omega_2$.
\begin{align}
    \overline{\pi_i \pi_j} = \overline{\pi_{i1}\pi_{j1}}+ \overline{\pi_{i2}\pi_{j2}}
\end{align}

Now, we can use  Eq.~\eqref{eq:classicalstress} to sum the two contributions to the time averaged stress and write down the non-zero terms to order $k^2l^2.$
\begin{align*}
   & \begin{pmatrix}
        \overline{T}^{x}_{~x}\\  \overline{T}^{y}_{~y}\\  \overline{T}^{z}_{~z}\\  \overline{T}^{x}_{~z}
    \end{pmatrix}\\
    &= \begin{pmatrix}
        1  & 1 - k^2l^2 & 0 & k^2l^2 \\
        1 - k^2l^2 & 1 - 2 k^2l^2 & k^2l^2 & 2k^2 l^2\\
        0 & k^2l^2& 1 & 1 - k^2l^2\\
        k l^2 & k l^2 & - k l^2 & - kl^2
    \end{pmatrix} \begin{pmatrix}
        \frac{1}{2}m v_{x0}^2 \\
           \frac{1}{2}m v_{y0}^2\\
              \frac{1}{2}m v_{z0}^2\\
                 \frac{\omega_0^2}{2}m z_0^2
    \end{pmatrix}
\end{align*}

Summing over many possible configurations weighted by the appropriate Boltzmann factor is equivalent to using the equipartition theorem which, as we have shown previously, returns an isotropic stress tensor. 
We may directly see the compensatory nature of the two modes (as discussed in section \ref{sec:EffMet}) in the above matrix where rows sum to 2 for all $k$ and $l$ except for the strain term $T^{x}_{~z}$, which sums to zero as expected.

\section{Calculation of Stress under the Approximation  $\langle Y^\dag Y\rangle_{\mathrm{th}} =0$}
\label{app:Analytic}

\begin{widetext}
While an analytic formula for the stress as a function of temperature would be useful, the sum over both integer modes is intractable in general. However, if we neglect the sum over the quasi-confining potential level $\kappa$ by using the approximation $\langle Y^\dag Y \rangle_{\mathrm{th}}=\kappa  = 0$ which is good when $T \ll \omega_2$, then the analytic formulation becomes calculable.
The approximated thermally weighted quantum stress is 

\begin{align}\label{eq:stressstartingpoint}
    \langle T^{\mu}_{~\nu}\rangle_{\mathrm{th}}\approx  \sum_{n =0 } T^{\mu}_{~\nu}(n,\kappa = 0 ) f(n,\kappa=0),
\end{align}
where $f(n,\kappa)$ is the Fermi-Dirac distribution.

First, we can use the Poisson summation formula of Eq.~\eqref{eq:Poisson} to Fourier transform Eq.~\eqref{eq:stressstartingpoint} as 
\begin{align}\label{eq:PoissonSum}
     &\langle T^{\mu}_{~\nu}\rangle_{\mathrm{th}} 
      = - \frac{1}{2}\frac{T^{\mu}_{~\nu}(0,0)}{1 + \exp(\beta (\omega_1/2+\omega_2/2-\mu))}+\sum_{q\in \mathbb Z } \int_{0}^\infty dx \frac{T^{\mu}_{~\nu}(x,0)}{1 + \exp(\beta (\omega_1(x+1/2) + \omega_2/2 -\mu))}e^{i 2\pi q x},
\end{align}
where $T^{x}_{~x}(x,0) =  \omega_1 (2x+1)$ and $T^{y}_{~y}(x,0) = \omega_1( 2x+1) (1 - k^2l^2) + k^2l$. 
Therefore we can evaluate the integral in Eq.~\eqref{eq:PoissonSum} for all components of the stress by considering arbitrary $T(x,0)= b$ and taking a derivative with respect to $q$ to deduce the first moment. 
We shift $\mu\xrightarrow[]{}\mu  +\omega_1/2+\omega_2/2$ to get rid of the constant terms.

Consider a long rectangular path in the upper-half complex plane bounded by the coordinates $z \in \lim_{x\xrightarrow[]{}\infty}\{ 0+i0, x+i0,x+ i \frac{2\pi}{\omega_1\beta},  0+i \frac{2\pi}{\omega_1\beta}\}$.
The rectangle encloses a single pole of the integrand at $z= \mu/\omega_1 + i \frac{\pi}{\beta\omega_1}$.
Therefore, 
\begin{align}
\begin{split}
     \int_{0}^\infty \frac{dx e^{i 2 \pi q x}}{1 + \exp(\beta (\omega_1 x-\mu))} -     \int_{0}^\infty \frac{dx e^{i 2 \pi q (x+ 2 \pi i/\omega_1\beta  )}}{1 + \exp(\beta  (\omega_1x+ 2 \pi i/\beta  - \mu ))} -     \int_{0}^{2\pi/\omega_1\beta } \frac{i dz e^{- 2 \pi q   z}}{1 + \exp(\beta \omega_1 (- \mu /\omega_1 + i z))}\\
    = 2\pi i \text{Res}\bigg[\frac{e^{i2\pi q x }}{1 + \exp(\beta(\omega_1 x-\mu))}\bigg].
    \end{split}
\end{align}

The horizontal legs are equal up to a multiplicative factor, and the residue at the simple pole can be found straightforwardly, yielding
\begin{align}
\begin{split}
      (1-  e^{- 4\pi^2q/\omega_1\beta}) \int_{0}^\infty \frac{dx e^{i 2 \pi q x}}{1 + \exp(\beta (\omega_1 x-\mu))}  -     \int_{0}^{2\pi/\omega_1\beta } \frac{i dz e^{- 2 \pi q   z}}{1 + \exp(\beta \omega_1 (- \mu /\omega_1 + i z))}
    =- \frac{2\pi ie^{- 2\pi^2 q /\beta \omega_1 + i 2\pi q \mu/\omega_1}}{\beta\omega_1}
    \end{split}.
\end{align}

The remaining integral along vertical leg may be calculated by Taylor expanding the integrand in powers of the inverse fugacity $e^{-\beta\mu}$.
\begin{align}
    \int_0^{2\pi/\omega_1\beta} \frac{i dz e^{-2\pi q z}}{1 + \exp(- \mu\beta)\exp(i\beta \omega_1 z)} =i \sum_{r=0} \left(-\exp(- \mu\beta )\right)^r  \int_0^{2\pi/\omega_1\beta} dz\exp\big((i  r\beta \omega_1 - 2 \pi q )z\big)
\end{align}

Putting the pieces together, we find that our original integral becomes
\begin{align}
    \int_0^\infty &\frac{dxe^{i 2 \pi x q }}{1 + \exp(\beta(\omega_1 x - \mu))} 
    \overset{q\neq 0}{=} -\frac{1}{\beta \omega_1 }\sum_{r = 0} \frac{(-e^{-\beta\mu})^r}{r +i 2 \pi q/\beta \omega_1 }-  \frac{i 2 \pi e^{ i 2 \pi q \mu/\omega_1}}{\beta \omega_1\sinh(2 \pi^2q/\beta \omega_1)}.
\end{align}

The $q = 0$ case may be solve exactly without resorting to a contour integral.
\begin{align}
    \int_0^\infty dx\frac{(2a x + b)}{1  + \exp(\beta(\omega_1 x - \mu))} = \frac{b}{\beta\omega_1}\ln(1 + e^{\beta \mu})- \frac{2a}{\beta^2\omega_1^2} \text{Li}_2(- e^{\beta\mu})
\end{align}

Where Li$_2$ is the second order polylogarithm function.
We combine the sum over $q$ and $r$ to a sum over $n$:
\label{eq:AnalyticEQ}
\begin{align}
\begin{split}
  \sum_{q=1}^\infty \int_0^\infty  dx\frac{(2a x + b)2 \cos(2\pi q x)}{1 + \exp(\beta(\omega_1 x - \mu))}=
    - \frac{a}{6} + &\sum_{n=1} \bigg[ \frac{(-e^{-\beta\mu})^n}{(8 n^2 \lambda^2)}\left(-4a + 8 bn^2\lambda^2 \text{coth}(n \lambda) + a n^2\beta^2\omega_1^2 \text{csch}^2(n \lambda) \right)\\
    & ~~~~~~~~~~~~~~+ \frac{ \pi b\sin(n\gamma) }{\lambda \sinh \frac{ \pi^2n }{\lambda}}+ \frac{\pi a }{\lambda ^2}  \text{csch}\frac{\pi^2n }{\lambda}\bigg( \pi \cos(n\gamma)  \text{coth}\frac{ \pi^2 n}{\lambda}+ \beta \mu \sin(n\gamma)\bigg)\bigg].
    \end{split}
\end{align}

Where $\lambda = \frac{\beta\omega_1}{2}$ and $\gamma = \frac{2\pi \mu}{\omega_1}$. Each of these components can be plugged into equation \eqref{eq:PoissonSum} to derive the expressions used in calculating the analytic function of figure \ref{fig:TempDepen}.
To recover $\langle T^{x}_{~x}\rangle_{\mathrm{th}}$  take $(a,b) =( \omega_1,\omega_1)$, and for $\langle T^{y}_{~y}\rangle_{\mathrm{th}}$ take $(a,b)=( \omega_1(1 - k^2l^2),\omega_1(1 - k^2l^2) + \omega_2 k^2l^2)$.

\end{widetext}

\end{appendix}

\bibliography{refs.bib}

\end{document}